# Deep Impact: Unintended consequences of journal rank

Björn Brembs[1], Katherine Button[2] and Marcus Munafò[3]

1. Institute of Zoology – Neurogenetics, University of Regensburg,

Universitätsstr. 31, 93040 Regensburg, Germany, bjoern@brembs.net

2. School of Social and Community Medicine, University of Bristol, 12a

Priory Road, Bristol BS8 1TU, United Kingdom.

3. UK Centre for Tobacco Control Studies and School of Experimental

Psychology, University of Bristol, 12a Priory Road, Bristol BS8 1TU,

United Kingdom.

Corresponding Author: Björn Brembs

## Abstract

Most researchers acknowledge an intrinsic hierarchy in the scholarly journals ('journal rank') that they submit their work to, and adjust not only their submission but also their reading strategies accordingly. On the other hand, much has been written about the negative effects of institutionalizing journal rank as an impact measure. So far, contributions to the debate concerning the limitations of journal rank as a scientific impact assessment tool have either lacked data, or relied on only a few studies. In this review, we present the most recent and pertinent data on the consequences of our current scholarly communication system with respect to various measures of scientific quality (such as utility/citations, methodological soundness, expert ratings or retractions). These data corroborate previous hypotheses: using journal rank as an assessment tool is bad scientific practice. Moreover, the data lead us to argue that any journal rank (not only the currently-favored Impact Factor) would have this negative impact. Therefore, we suggest that abandoning journals altogether, in favor of a library-based scholarly communication system, will ultimately be necessary. This new system will use modern information technology to vastly improve the filter, sort and discovery functions of the current journal system.

# Introduction

Science is the bedrock of modern society, improving our lives through advances in medicine, communication, transportation, forensics, entertainment and countless other areas. Moreover, today's global problems cannot be solved without scientific input and understanding. The more our society relies on science, and the more our population becomes scientifically literate, the more important the reliability (i.e., veracity and integrity, or, 'credibility' (Ioannidis, 2012)) of scientific research becomes. Scientific research is largely a public endeavor, requiring public trust. Therefore, it is critical that public trust in science remains high. In other words, the reliability of science is not only a societal imperative, it is also vital to the scientific community itself. However, every scientific publication may in principle report results which prove to be unreliable, either unintentionally, in the case of honest error or statistical variability, or intentionally in the case of misconduct or fraud. Even under ideal circumstances, science can never provide us with absolute truth. In Karl Popper's words: "Science is not a system of certain, or established, statements" (Popper, 1995). Peer-review is one of the mechanisms which have evolved to increase the reliability of the scientific literature.

At the same time, the current publication system is being used to structure the careers of the members of the scientific community by evaluating their success in obtaining publications in high-ranking journals. The hierarchical publication system ('journal rank') used to communicate scientific results is thus central, not only to the composition of the scientific community at large (by selecting its members), but also to science's position in society. In recent years, the scientific study of the effectiveness of such measures of quality control has grown.

## Retractions and the Decline Effect

A disturbing trend has recently gained wide public attention: The retraction rate of articles published in scientific journals, which had remained stable since the 1970's, began to increase rapidly in the early 2000's from 0.001% of the total to about 0.02% (Figure 1a). In 2010 we have seen the creation and popularization of a website dedicated to monitoring retractions (http://retractionwatch.com), while 2011 has been described as the "the year of the retraction" (Hamilton, 2011). The reasons suggested for retractions vary widely, with the recent sharp rise potentially facilitated by an increased willingness of journals to issue retractions, or increased scrutiny and error-detection from online media. Although cases of clear scientific misconduct initially constituted a minority of cases (Fanelli, 2009; Van Noorden, 2011; Wager and Williams, 2011; Nath et al., 2006; Cokol et al., 2007; Steen, 2011a), the fraction of retractions due to misconduct has risen sharper than the overall retraction rate and now the majority of all retractions is due to misconduct (Fang et al., 2012; Steen, 2011b).

Retraction notices, a metric which is relatively easy to collect, only constitute the extreme end of a spectrum of unreliability that is inherent to the scientific method: we can hardly ever be *entirely* certain of our results (Popper, 1995). Much of the training scientists receive aims to reduce this uncertainty long before the work is submitted for publication. However, a less readily quantified but more frequent phenomenon (compared to rare retractions) has recently garnered attention, which calls into question the effectiveness of this training. The 'decline-effect', which is now well-described, relates to the observation that the strength of evidence for a particular finding often declines over time (Schooler, 2011; Lehrer, 2010; Bertamini and Munafo, 2012; Palmer, 2000; Fanelli, 2010; Ioannidis, 2005b; Simmons et al., 1999, 2011; Møller and Jennions, 2001; Møller et al., 2005; Van Dongen, 2011; Gonon et al., 2012). This effect provides wider scope for assessing the unreliability of scientific research than retractions alone, and allows for more general conclusions to be drawn.

Researchers make choices about data collection and analysis which increase the chance of false-positives (i.e., researcher bias) (Simmons et al., 1999, 2011), and surprising and novel effects are more likely to be published than studies showing no effect. This is the well-known phenomenon of publication bias (Song et al., 1999; Van Dongen, 2011; Munafò et al., 2007; Young et al., 2008; Callaham, 2002; Møller and Jennions, 2001; Møller et al., 2005; Schooler, 2011; Dwan et al., 2008). In other words, the probability of getting a paper published might be biased towards larger initial effect sizes, which are revealed by later studies to be not so large (or even absent entirely), leading to the decline effect. While sound methodology can help reduce researcher bias (Simmons et al., 1999), publication bias is more difficult to address. Some journals are devoted to publishing null results, or have sections devoted to these, but coverage is uneven across disciplines and often these are not particularly high-ranking or well-read (Schooler, 2011; Nosek et al., 2012). Publication therein is typically not a cause for excitement (Giner-Sorolla, 2012; Nosek et al., 2012), leading to an overall low frequency of replication studies in many fields (Hartshorne and Schachner, 2012; Kelly, 2006; Carpenter, 2012; Yong, 2012; Makel et al., 2012). Publication bias is also exacerbated by a tendency for journals to be less likely to publish replication studies (or, worse still, failures to replicate) (Editorial, 2012; Goldacre, 2011; Sutton, 2011; Hartshorne and Schachner, 2012; Curry, 2009; Yong, 2012). Here we argue that the counter-measures proposed to improve the reliability and veracity of science such as peer-review in a hierarchy of journals or methodological training of scientists may not be sufficient.

While there is growing concern regarding the increasing rate of retractions in particular, and the unreliability of scientific findings in general, little consideration has been given to the infrastructure by which scientists not only communicate their findings but also evaluate each other as a potential contributing factor. That is, to what extent does the environment in which science takes place contribute to the problems described above? By far the most common metric by which publications are evaluated, at least initially, is the perceived prestige or rank of the journal in which they appear. Does the pressure to publish in prestigious, high-ranking journals contribute to the unreliability of science?

## The Decline Effect and Journal Rank

The common pattern seen where the decline effect has been documented is one of an initial publication in a high-ranking journal, followed by attempts at replication in lower-ranked journals which either failed to replicate the original findings, or suggested a much weaker effect (Lehrer, 2010). Journal rank is most commonly assessed using Thomson Reuters' Impact Factor (IF), which has been shown to correspond well with subjective ratings of journal quality and rank (Gordon, 1982; Saha et al., 2003; Yue et al., 2007; Sønderstrup-Andersen and Sønderstrup-Andersen, 2008). One particular case (Munafò et al., 2007) illustrates the decline effect (Figure 1b), and shows that early publications both report a larger effect than subsequent studies, and are also published in journals with a higher IF. These observations raise the more general question of whether research published in high-ranking journals is inherently less reliable than research in lower-ranking journals.

As journal rank is also predictive of the incidence of fraud and misconduct in retracted publications, as opposed to other reasons for retraction (Steen, 2011a), it is not surprising that higher ranking journals are also more likely to publish fraudulent work than lower ranking journals (Fang et al., 2012). These data, however, cover only the small fraction of publications that have been retracted. More important is the large body of the literature that is not retracted and thus actively being used by the scientific community. There is evidence that unreliability is higher in high-ranking journals as well, also for non-retracted publications: A meta-analysis of genetic association studies provides evidence that the extent to which a study over-estimates the likely true effect size is positively correlated with the IF of the journal in which it is published (Figure 1c) (Munafò et al., 2009). Similar effects have been reported in the context of other research fields (Siontis et al., 2011; Ioannidis, 2005a; Ioannidis and Panagiotou, 2011).

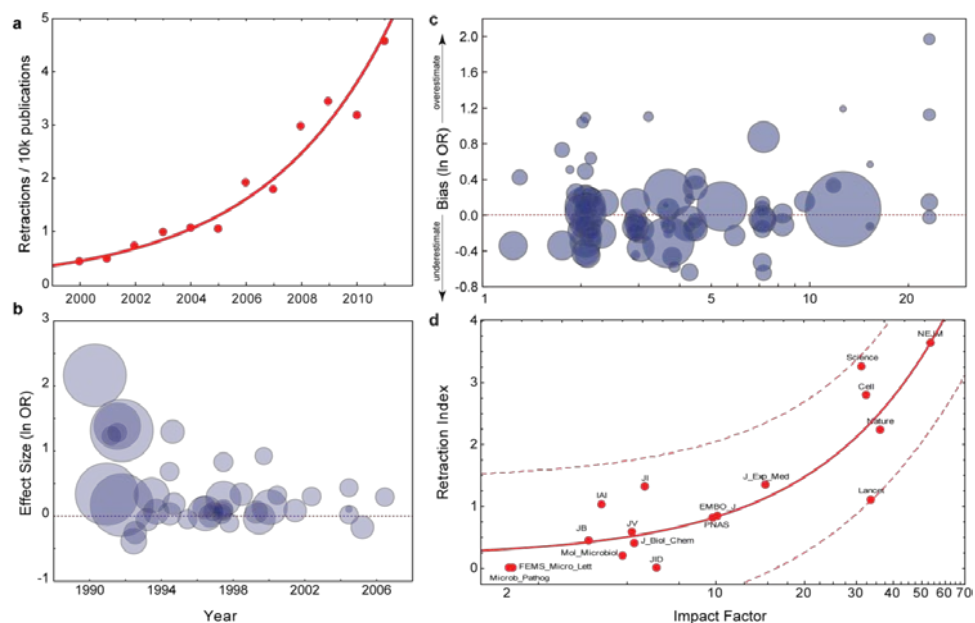

**Fig. 1:** *Current trends in the reliability of science.*
**a** – Exponential fit for PubMed retraction notices (data from pmretract.heroku.com). **b** – Relationship between year of publication and individual study effect size. Data are taken from Munafò et al., 2007, and represent candidate gene studies of the association between *DRD2* genotype and alcoholism. The effect size (y-axis) represents the individual study effect size (odds ratio; OR), on a log-scale. This is plotted against the year of publication of the study (x-axis). The size of the circle is proportional to the IF of the journal the individual study was published in. Effect size is significantly negatively correlated with year of publication. **c** – Relationship between IF and extent to which an individual study overestimates the likely true effect. Data are taken from Munafò et al., 2009, and represent candidate gene studies of a number of gene-phenotype associations of psychiatric phenotypes. The bias score (y-axis) represents the effect size of the individual study divided by the pooled effect size estimated indicated by meta-analysis, on a log-scale. Therefore, a value greater than zero indicates that the study provided an over-estimate of the likely true effect size. This is plotted against the IF of the journal the study was published in (x-axis), on a log-scale. The size of the circle is proportional to the sample size of the individual study. Bias score is significantly positively correlated with IF, sample size significantly negatively. **d** – Linear regression with confidence intervals between IF and Fang and Casadevall's Retraction Index (data provided by Fang and Casadevall, 2011).

There are additional measures of scientific quality and in none does journal rank fare much better. A study in crystallography reports that the quality of the protein structures described is significantly lower in publications in high-ranking journals (Brown and Ramaswamy, 2007). Adherence to basic principles of sound scientific (e.g., the CONSORT guidelines: http://www.consort-statement.org), or statistical methodology have also been tested. Four different studies on levels of evidence in medical and/or psychological research have found varying results. While two studies on surgery journals found a correlation between IF and the levels of evidence defined in the respective studies (Obremskey et al., 2005; Lau and Samman, 2007), a study of anesthesia journals failed to find any statistically significant correlation between journal rank and evidence-based medicine principles (Bain and Myles, 2005) and a study of seven medical/psychological journals found highly varying adherence to statistical guidelines, irrespective of journal rank (Tressoldi et al., 2013). The two surgery studies covered an IF range between 0.5 and 2.0, and 0.7 and 1.2, respectively, while the anesthesia study covered the range 0.8 to 3.5. It is possible that any correlation at the lower end of the scale is abolished when higher rank journals are included. The study by Tressoldi and colleagues, which included very high ranking journals, supports this interpretation. Importantly, if publications in higher ranking journals were methodologically sounder, then one would expect the opposite result: inclusion of high-ranking journals should result in a stronger, not a weaker correlation. Further supporting the notion that journal rank is a poor predictor of statistical soundness is our own analysis of data on statistical power in neuroscience studies (Button et al., 2013). There was no significant correlation between statistical power and journal rank (N=650; $r_s$=-0.01; t=0.8; Figure 2). Thus, the currently available data seem to indicate that journal rank is a poor indicator of methodological soundness.

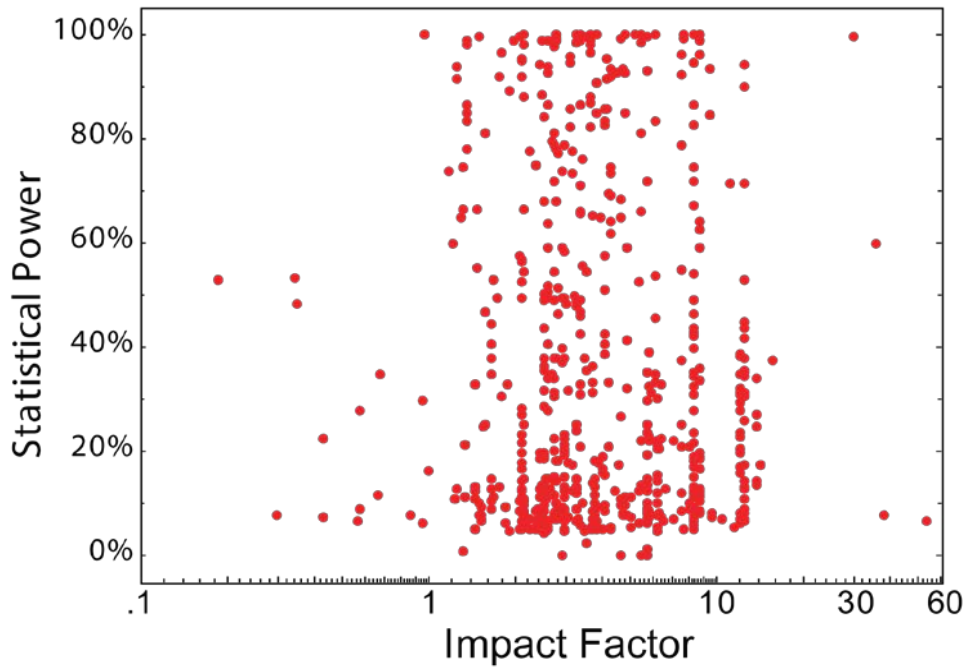

**Fig. 2:** *No association between statistical power and journal IF.*
The statistical power of 650 neuroscience studies (data from Button et al. 2013; 19 missing ref; 3 unclear reporting; 57 published in journal without 2011 IF; 1 book) plotted as a function of the 2011 IF of the publishing journal. The studies were selected from the 730 contributing to the meta-analyses included in Button et al. 2013, Table 1, and included where journal title and IF (2011 © Thomson Reuters Journal Citation Reports) were available.

Beyond explicit quality metrics and sound methodology, reproducibility is at the core of the scientific method and thus a hallmark of scientific quality. Three recent studies reported attempts to replicate published findings in preclinical medicine (Scott et al., 2008; Begley and Ellis, 2012; Prinz et al., 2011). All three found a very low frequency of replication, suggesting that maybe only one out of five preclinical findings is reproducible. In fact, the level of reproducibility was so low that no relationship between journal rank and reproducibility could be detected. Hence, these data support the necessity of recent efforts such as the 'Reproducibility Initiative' (Baker, 2012) or the "Reproducibility Project" (Collaboration, 2012) . In fact, the data also indicate that these projects may consider starting with replicating findings published in high-ranking journals.

Given all of the above evidence, it is therefore not surprising that journal rank is also a strong predictor of the rate of retractions (Figure 1d) (Fang and Casadevall, 2011; Liu, 2006; Cokol et al., 2007).

# Social pressure and journal rank

There are thus several converging lines of evidence which indicate that publications in high ranking journals are not only more likely to be fraudulent than articles in lower ranking journals, but also more likely to present discoveries which are less reliable (i.e., are inflated, or cannot subsequently be replicated). Some of the sociological mechanisms behind these correlations have been documented, such as pressure to publish (preferably positive results in high-ranking journals), leading to the potential for decreased ethical standards (Anderson et al., 2007) and increased publication bias in highly competitive fields (Fanelli, 2010). The general increase in competitiveness, and the precariousness of scientific careers (Shapin, 2008), may also lead to an increased publication bias across the sciences (Fanelli, 2011). This evidence supports earlier propositions about social pressure being a major factor driving misconduct and publication bias (Giles, 2007), eventually culminating in retractions in the most extreme cases.

That being said, it is clear that the correlation between journal rank and retraction rate is likely too strong (coefficient of determination of 0.77; data from (Fang and Casadevall, 2011)) to be explained exclusively by the decreased reliability of the research published in high ranking journals. Probably, additional factors contribute to this effect. For instance, one such factor may be the greater visibility of publications in these journals, which is both one of the incentives driving publication bias, and a likely underlying cause for the detection of error or misconduct with the eventual retraction of the publications as a result (Cokol et al., 2007). Conversely, the scientific community may also be less concerned about incorrect findings published in more obscure journals. With respect to the latter, the finding that the large majority of retractions come from the numerous lower-ranking journals (Fang et al., 2012) reveals that publications in lower ranking journals are scrutinized and, if warranted, retracted. Thus, differences in scrutiny are likely to be only a contributing factor and not an exclusive explanation, either. With respect to the former, visibility effects in general can be quantified by measuring citation rates between journals, testing the assumption that if higher visibility were a contributing factor to retractions, it must also contribute to citations.

## Journal Rank and Study Impact

Thus far we have presented evidence that research published in high-ranking journals may be less reliable compared with publications in lower-ranking journals. Nevertheless, there is a strong common perception that high-ranking journals publish 'better' or 'more important' science, and that the IF captures this well (Gordon, 1982; Saha et al., 2003). The assumption is that high-ranking journals are able to be highly selective and publish only the most important, novel and best-supported scientific discoveries, which will then, as a consequence of their quality, go on to be highly cited (Young et al., 2008). One way to reconcile this common perception with the data would be that, while journal rank may be indicative of a minority of unreliable publications, it may also (or more strongly) be indicative of the importance of the majority of remaining, reliable publications. Indeed, a recent study on clinical trial meta-analyses found that a measure for the novelty of a clinical trial's main outcome did correlate significantly with journal rank (Evangelou et al., 2012). Compared to this relatively weak correlation (with all coefficients of determination lower than 0.1), a stronger correlation was reported for journal rank and expert ratings of importance (Allen et al., 2009). In this study, the journal in which the study had appeared was not masked, thus not excluding the strong correlation between subjective journal rank and journal quality as a confounding factor. Nevertheless, there is converging evidence from two studies that journal rank is indeed indicative of a publication's perceived importance.

Beyond the importance or novelty of the research, there are three additional reasons why publications in high-ranking journals might receive a high number of citations. First, publications in high-ranking journals achieve greater exposure by virtue not only of the larger circulation of the journal in which they appear, but also of the more prominent media attention (Gonon et al., 2012). Second, citing high-ranking publications in one's own publication may increase its perceived value. Third, the novel, surprising, counter-intuitive or controversial findings often published in high-ranking journals, draw citations not only from follow-up studies but also from news-type articles in scholarly journals reporting and discussing the discovery. Despite these four factors, which would suggest considerable effects of journal rank on future citations, it has been established for some time that the actual effect of journal rank is measurable, but nowhere near as substantial as indicated (Hegarty and Walton, 2012; Seglen, 1997; Callaham, 2002; Kravitz and Baker, 2011; Chow et al., 2007; Seglen, 1994; Finardi, 2013) and as one would expect if visibility were the exclusive factor driving retractions. In fact, the average effect sizes roughly approach those for journal rank and unreliability, cited above.

The data presented in a recent analysis of the development of these correlations between IF-based journal rank and future citations over the period from 1902-2009 (with IFs before the 1960s computed retroactively) reveal two very informative trends (Figure 3, data from (Lozano et al., 2012). First, while the

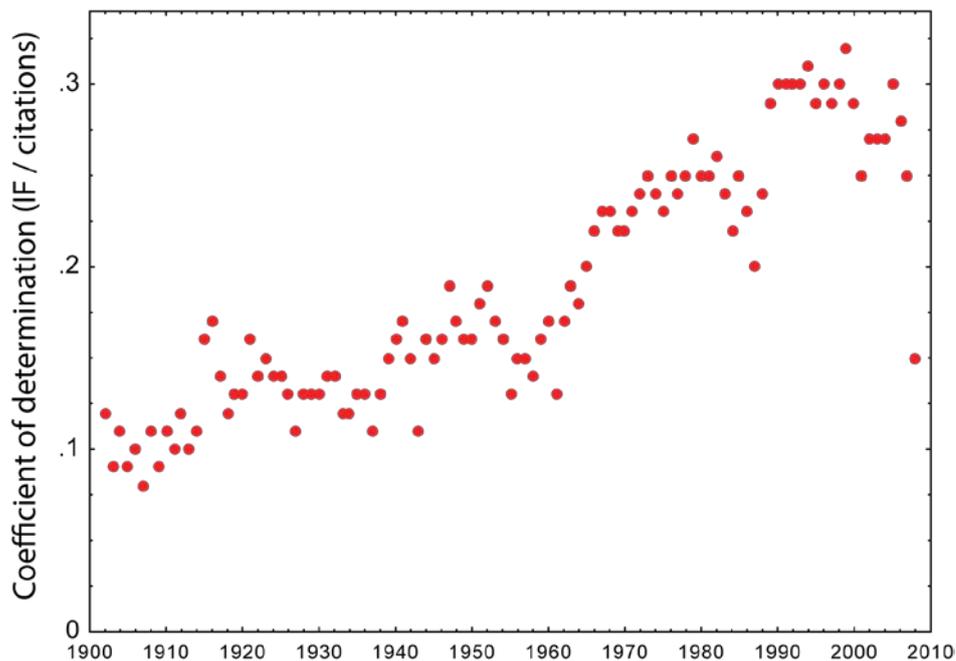

**Fig. 3:** *Trends in predicting citations from journal rank.*

The coefficient of determination ($R^2$) between journal rank (as measured by IF) and the citations accruing over two years after publications is plotted as a function of publication year in a sample of almost 30 million publications. Lozano et al. (2012) make the case that one can explain the trends in the predictive value of journal rank by the publication of the IF in the 1960s ($R^2$ increase is accelerating) and the widespread adoption of internet searches in the 1990s ($R^2$ is dropping). The data support the interpretation that reading habits drive the correlation between journal rank and citations more than any inherent quality of the articles. IFs before the invention of the IF have been retroactively computed for the years before the 1960s.

predictive power of journal rank remained very low for the entire first two thirds of the 20th century, it started to slowly increase shortly after the publication of the first IF data in the 1960's. This correlation kept increasing until the second interesting trend emerged with the advent of the internet and keyword-search engines in the 1990's, from which time on it fell back to pre-1960's levels until the end of the study period in 2009. Overall, consistent with the citation data already available, the coefficient of determination between journal rank and citations was always in the range of ~0.1 to 0.3 (i.e., quite low). It thus appears that indeed a small but significant correlation between journal rank and future citations can be observed. Moreover, the data suggest that most of this small effect stems from visibility effects due to the influence of the IF on reading habits (Lozano et al., 2012), rather than from factors intrinsic to the published articles (see data cited above). However, the correlation is so weak that it cannot alone account for the strong correlation between retractions and journal rank, but instead requires additional factors, such as the increased unreliability of publications in high ranking journals cited above. Supporting these weak correlations between journal rank and future citations are data reporting classification errors (i.e., whether a publication received too many or too few citations with regard to the rank of the journal it was published in) at or exceeding 30% (Chow et al., 2007; Kravitz and Baker, 2011; Singh et al., 2007; Starbuck, 2005). In fact, these classification errors, in conjunction with the weak citation advantage, render journal rank practically useless as an evaluation signal, even if there was no indication of less reliable science being published in high ranking journals.

The only measure of citation count that does correlate strongly with journal rank (negatively) is the number of articles without any citations at all (Weale et al., 2004), supporting the argument that fewer articles in high-ranking journals go unread. Thus, there is quite extensive evidence arguing for the strong correlation between journal rank and retraction rate to be mainly due to two factors: there is direct evidence that the social pressures to publish in high ranking journals increases the unreliability, intentional or not, of the research published there. There is more indirect evidence, derived mainly from citation data, indicating that increased visibility of publications in high ranking journals may potentially contribute to increased error-detection in these journals. With several independent measures failing to provide compelling evidence that journal rank is a reliable predictor of scientific impact or quality, and other measures indicating that journal rank is at least equally if not more predictive of low reliability, the central role of journal rank in modern science deserves close scrutiny.

# Practical consequences of Journal Rank

Even if a particular study has been performed to the highest standards, the quest for publication in high-ranking journals slows down the dissemination of science and increases the burden on reviewers, by iterations of submissions and rejections cascading down the hierarchy of journal rank (Statzner and Resh, 2010; Kravitz and Baker, 2011; Nosek and Bar-Anan, 2012). A recent study seems to suggest that such rejections eventually improve manuscripts enough to yield measurable citation benefits (Calcagno et al., 2012). However, the effect size of such resubmissions appears to be of the order of 0.1 citations per article, a statistically significant but, in practical terms, negligible effect. This conclusion is corroborated by an earlier study which failed to find any such effect (Nosek and Bar-Anan, 2012). Moreover, with peer-review costs estimated in excess of 2.2 billion € (US$~2.8b) annually (Research Information Network, 2008), the resubmission cascade contributes to the already rising costs of journal rank: the focus on journal rank has allowed corporate publishers to keep their most prestigious journals closed-access and to increase subscription prices (Kyrillidou et al., 2012), creating additional barriers to the dissemination of science. The argument from highly selective journals is that their per-article cost would be too high for author processing fees, which may be up to 37,000€ (US$48,000) for the journal *Nature* (House of Commons, 2004). There is also evidence from one study in economics suggesting that journal rank can contribute to suppression of interdisciplinary research (Rafols et al., 2012), keeping disciplines separate and isolated.

Finally, the attention given to publication in high-ranking journals may distort the communication of scientific progress, both inside and outside of the scientific community. For instance, the recent discovery of a 'Default-Mode Network' in rodent brains was, presumably, made independently by two different sets of neuroscientists and published only a few months apart (Lu et al., 2012; Upadhyay et al., 2011). The later, but not the earlier, publication (Lu et al., 2012) was cited in a subsequent high-ranking publication (Welberg, 2012). Despite both studies largely reporting identical findings (albeit, perhaps, with different quality), the later report has garnered 19 citations, while the earlier one only 5, at the time of this writing. We do not know of any empirical studies quantitatively addressing this particular effect of journal rank. However, a similar distortion due to selective attention to publications in high-ranking journals has been reported in a study on medical research. This study found media reporting to be distorted, such that once initial findings in higher-ranking journals have been refuted by publications in lower ranking journals (a case of decline effect), they do not receive adequate media coverage (Gonon et al., 2012).

## Impact Factor – Negotiated, irreproducible and unsound

The IF is a metric for the number of citations to articles in a journal (the numerator), normalized by the number of articles in that journal (the denominator). However, there is evidence that IF is, at least in some cases, not calculated but negotiated, that it is not reproducible, and that, even if it were reproducibly computed, the way it is derived is not mathematically sound. The fact that publishers have the option to negotiate how their IF is calculated is well-established – in the case of *PLoS Medicine*, the negotiation range was between 2 and about 11 (The PLoS Medicine Editors, 2006). What is negotiated is the denominator in the IF equation (i.e., which published articles which are counted), given that all citations count towards the numerator whether they result from publications included in the denominator or not. It has thus been public knowledge for quite some time now that removing editorials and News-and-Views articles from the denominator (so called "front-matter") can dramatically alter the resulting IF (Editorial, 2005; Garfield, 1999; Adam, 2002; Moed and Van Leeuwen, 1995; Moed and van Leeuwen, 1996; Hernán, 2009; Baylis et al., 1999). While these IF negotiations are rarely made public, the number of citations (numerator) and published articles (denominator) used to calculate IF are accessible via *Journal Citation Reports*. This database can be searched for evidence that the IF has been negotiated. For instance, the numerator and denominator values for *Current Biology* in 2002 and 2003 indicate that while the number of citations remained relatively constant, the number of published articles dropped. This decrease occurred after the journal was purchased by Cell Press (an imprint of Elsevier), despite there being no change in the layout of the journal. Critically, the arrival of a new publisher corresponded with a *retrospective* change in the denominator used to calculate IF (Table 1). Similar procedures raised the IF of *FASEB Journal* from 0.24 in 1988 to 18.3 in 1989, when conference abstracts ceased to count towards the denominator (Baylis et al., 1999).

| Journal: Current Biology | Published items 2000 | Published items 2001 | Published items 2002 | Sum published items | Citations in preceding two years | IF |
|---|---|---|---|---|---|---|
| JCR Science Edition 2002 | 504 | 528 | n.c. | 1032 | 7231 | 7.007 |
| JCR Science Edition 2003 | n.c. | 300 | 334 | 634 | 7551 | 11.910 |

**Table 1:** Thomson Reuters' IF calculations for the journal 'Current Biology' in the years 2002/2003.
Most of the rise in IF is due to the reduction in published items. Note the discrepancy between the number of items published in 2001 between the two consecutive JCR Science Editions. – n.c.: year not covered by this edition. Raw data see Suppl. Fig. S1.

In an attempt to test the accuracy of the ranking of some of their journals by IF, Rockefeller University Press purchased access to the citation data of their journals and some competitors. They found numerous discrepancies between the data they received and the published rankings, sometimes leading to differences of up to 19% (Rossner et al., 2007). When asked to explain this discrepancy, Thomson Reuters replied that they routinely use several different databases and had accidentally sent Rockefeller University Press the wrong one. Despite this, a second database sent also did not match the published records. This is only one of a number reported errors and inconsistencies (Reedijk, 1998; Moed et al., 1996).

It is well-known that citation data are strongly left-skewed, meaning that a small number of publications receive a large number of citations, while most publications receive very few (Rossner et al., 2007; Seglen, 1992, 1997; Kravitz and Baker, 2011; Editorial, 2005; Chow et al., 2007; Weale et al., 2004; Taylor et al., 2008). The use of an arithmetic mean as a measure of central tendency on such data (rather than, say, the median) is clearly inappropriate, but this is exactly what is used in the IF calculation. The International Mathematical Union reached the same conclusion in an analysis of the IF (Adler et al., 2008). A recent study correlated the median citation frequency in a sample of 100 journals with their two-year IF and found a very strong correlation, which is expected due to the similarly left-skewed distributions in most journals (Editorial, 2013). However, at the time of this writing, it is not known if using the median (instead of the mean) improves any of the predominantly weak predictive properties of journal rank. Complementing the specific flaws just mentioned, a recent, comprehensive review of the bibliometric literature lists various additional shortcomings of the IF more generally (Vanclay, 2011).

## Conclusions

While at this point it seems impossible to quantify the relative contributions of the different factors influencing the reliability of scientific publications, the current empirical literature on the effects of journal rank provides evidence supporting the following four conclusions: 1) journal rank is a weak to moderate predictor of utility and perceived importance; 2) journal rank is a moderate to strong predictor of both intentional and unintentional scientific unreliability; 3) journal rank is expensive, delays science and frustrates researchers; and, 4) journal rank as established by IF violates even the most basic scientific standards, but predicts subjective judgments of journal quality.

# Caveats

While our latter two conclusions appear uncontroversial, the former two are counter-intuitive and require explanation. Weak correlations between future citations and journal rank based on IF may be caused by the poor statistical properties of the IF. This explanation could (and should) be tested by using any of the existing alternative ranking tools available (such as Thomson Reuters' Eigenfactor, Scopus' SCImagoJournalRank, or Google's Scholar Metrics etc.) and computing correlations with the metrics discussed above. However, a recent analysis shows a high correlation between these ranks, so no large differences would be expected (Lopez-Cozar and Cabezas-Clavijo, 2013). Alternatively, one can choose other important metrics and compute which journals score particularly high on these. Either way, since the IF reflects the common perception of journal hierarchies rather well (Gordon, 1982; Saha et al., 2003; Yue et al., 2007; Sønderstrup-Andersen and Sønderstrup-Andersen, 2008), any alternative hierarchy that would better reflect article citation frequencies might violate this intuitive sense of journal rank, as different ways to compute journal rank lead to different hierarchies (Wagner, 2011). Both alternatives thus challenge our subjective journal ranking. To put it more bluntly, if perceived importance and utility were to be discounted as indirect proxies of quality, while retraction rate, replicability, effect size overestimation, correct sample sizes, crystallographic quality, sound methodology and so on counted as more direct measures of quality, then inversing the current IF-based journal hierarchy would improve the alignment of journal rank for most and have no effect on the rest of these more direct measures of quality.

The subjective journal hierarchy also leads to a circularity that confounds many empirical studies. That is, authors use journal rank, in part, to make decisions of where to submit their manuscripts, such that well-performed studies yielding ground-breaking discoveries with general implications are preferentially submitted to high-ranking journals. Readers, in turn, expect only to read about such articles in high-ranking journals, leading to the exposure and visibility confounds discussed above and at length in the cited literature. Moreover, citation practices and methodological standards vary in different scientific fields, potentially distorting both the citation and reliability data. Given these confounds one might expect highly varying and often inconclusive results. Despite this, the literature contains evidence for associations between journal rank and measures of scientific impact (e.g., citations, importance, unread articles), but also contains at least equally strong, consistent effects of journal rank predicting scientific unreliability (e.g., retractions, effect size, sample size, replicability, fraud/misconduct, methodology). Neither group of studies can thus be easily dismissed, suggesting that the incentives journal rank creates for the scientific community (to submit either their best or their most unreliable work to the most high-ranking journals) at best cancel each other out. Such unintended consequences are well-known from other fields where metrics are applied (Hauser and Katz, 1998).

Therefore, while there are concerns not only about the validity of the IF as the metric of choice for establishing journal rank but also about confounding factors complicating the interpretation of some of the data, we find, in the absence of additional data, that these concerns do not suffice to substantially question our conclusions, but do emphasize the need for future research.

# Potential long-term consequences of journal rank

Taken together, the reviewed literature suggests that using journal rank is unhelpful at best and unscientific at worst. In our view, IF generates an illusion of exclusivity and prestige based on an assumption that it will predict subsequent impact, which is not supported by empirical data. As the IF aligns well with intuitive notions of journal hierarchies (Gordon, 1982; Saha et al., 2003; Yue et al., 2007), it receives insufficient scrutiny (Frank, 2003) (perhaps a case of confirmation bias). The one field in which journal rank is scrutinized is bibliometrics. We have reviewed the pertinent empirical literature to supplement the largely argumentative discussion on the opinion pages of many learned journals (Adler and Harzing, 2009; Bauer, 2004; Lawrence, 2002; Brumback, 2012; Lawrence, 2007, 2008; Garwood, 2011; Taylor et al., 2008; Tsikliras, 2008; Todd and Ladle, 2008; Giles, 2007; Moed

and van Leeuwen, 1996; Editorial, 2005; Sarewitz, 2012; Schooler, 2011) with empirical data. Much like dowsing, homeopathy or astrology, journal rank seems to appeal to subjective impressions of certain effects, but these effects disappear as soon as they are subjected to scientific scrutiny.

In our understanding of the data, the social and psychological influences described above are, at least to some extent, generated by journal rank itself, which in turn may contribute to the observed decline effect and rise in retraction rate. That is, systemic pressures on the author, rather than increased scrutiny on the part of the reader, inflate the unreliability of much scientific research. Without reform of our publication system, the incentives associated with increased pressure to publish in high-ranking journals will continue to encourage scientists to be less cautious in their conclusions (or worse), in an attempt to market their research to the top journals (Anderson et al., 2007; Fanelli, 2010; Shapin, 2008; Giles, 2007; Munafò et al., 2009). This is reflected in the decline in null results reported across disciplines and countries (Fanelli, 2011), and corroborated by the findings that much of the increase in retractions may be due to misconduct (Steen, 2011b; Fang et al., 2012), and that much of this misconduct occurs in studies published high-ranking journals (Steen, 2011a; Fang et al., 2012). Inasmuch as journal rank guides the appointment and promotion policies of research institutions, the increasing rate of misconduct that has recently been observed may prove to be but the beginning of a pandemic: It is conceivable that, for the last few decades, research institutions world-wide may have been hiring and promoting scientists who excel at marketing their work to top journals, but who are not necessarily equally good at conducting their research. Conversely, these institutions may have purged excellent scientists from their ranks, whose marketing skills did not meet institutional requirements. If this interpretation of the data is correct, a generation of excellent marketers (possibly, but not necessarily, also excellent scientists) now serve as the leading figures and role models of the scientific enterprise, constituting another potentially major contributing factor to the rise in retractions.

The implications of the data presented here go beyond the reliability of scientific publications – public trust in science and scientists has been in decline for some time in many countries (Gauchat, 2010; EuropeanCommission, 2010; Nowotny, 2005), dramatically so in some sections of society (Gauchat, 2012), culminating in the sentiment that scientists are nothing more than yet another special interest group (Miller, 2012; Sarewitz, 2013). In the words of Daniel Sarewitz: "Nothing will corrode public trust more than a creeping awareness that scientists are unable to live up to the standards that they have set for themselves" (Sarewitz, 2012). The data presented here prompt the suspicion that the corrosion has already begun and that journal rank may have played a part in this decline as well.

## Alternatives

Alternatives to journal rank exist – we now have technology at our disposal which allows us to perform all of the functions journal rank is currently supposed to perform in an unbiased, dynamic way on a per-article basis, allowing the research community greater control over selection, filtering, and ranking of scientific information (Lin, 2012; Kravitz and Baker, 2011; Priem et al., 2012; Hönekopp and Khan, 2011; Roemer and Borchardt, 2012; Priem, 2013). Since there is no technological reason to continue using journal rank, one implication of the data reviewed here is that we can instead use current technology and remove the need for a journal hierarchy completely. As we have argued, it is not only technically obsolete, but also counter-productive and a potential threat to the scientific endeavor. We therefore would favor bringing scholarly communication back to the research institutions in an archival publication system in which both software, raw data and their text descriptions are archived and made accessible, after peer-review and with scientifically-tested metrics accruing reputation in a constantly improving reputation system (Eve, 2012). This reputation system would be subjected to the same standards of scientific scrutiny as are commonly applied to all scientific matters and evolve to minimize gaming and maximize the alignment of researchers' interests with those of science (which are currently misaligned (Nosek et al., 2012)). Only an elaborate ecosystem of a multitude of metrics can provide the flexibility to capitalize on the small fraction of the multi-faceted scientific output that is actually quantifiable. Such an ecosystem would evolve such that the only evolutionary stable strategy is to try and do the best science one can.

The currently balkanized literature, with a lack of interoperability and standards as one of its many detrimental, unintended consequences, prevents the kind of innovation that gave rise to the discover functions of Amazon or eBay, the social networking functions of Facebook or Reddit and course the sort and search functions of Google – all technologies virtually every scientist uses regularly for all activities but science. Thus, fragmentation and the resulting lack of access and interoperability are among the main underlying reasons why journal rank has not yet been replaced by more scientific evaluation options, despite widespread access to article-level metrics today. With an openly accessible scholarly literature standardized for interoperability, it would of course still be possible to pay professional editors to select publications, as is the case now, but after publication. These editors would then actually compete with each other for paying customers, accumulating track records for selecting (or missing) the most important discoveries. Likewise, virtually any functionality the current system offers would easily be replicable in the system we envisage. However, above and beyond replicating current functionality, an open, standardized scholarly literature would place any and all thinkable scientific metrics only a few lines of code away, offering the possibility of a truly open evaluation system where any hypothesis can be tested. Metrics, social networks and intelligent software then can provide each individual user with regular, customized updates on the most relevant research. These updates respond to the behavior of the user and learn from and evolve with their preferences. With openly accessible, interoperable literature, data and software, agents can be developed that independently search for hypotheses in the vast knowledge accumulating there. But perhaps most importantly, with an openly accessible database of science, innovation can thrive, bringing us features and ideas nobody can think of today and nobody will ever be capable of imagining, if we do not bring the products of our labor back under our own control. It was the hypertext transfer protocol (http) standard that spurred innovation and made the internet what it is today. What is required is the equivalent of http for scholarly literature, data and software.

Funds currently spent on journal subscriptions could easily suffice to finance the initial conversion of scholarly communication, even if only as long-term savings. One avenue to move in this direction may be the recently announced Episcience Project (Van Noorden, 2013). Other solutions certainly exist (Beverungen et al., 2012; Nosek and Bar-Anan, 2012; Kriegeskorte et al., 2012; Bachmann, 2011; Birukou et al., 2011; Florian, 2012; Ghosh et al., 2012; Hunter, 2012; Ietto-Gillies, 2012; Kreiman and Maunsell, 2011; Kriegeskorte, 2012; Lee, 2012; Pöschl, 2012; Priem and Hemminger, 2012; Sandewall, 2012; Walther and van den Bosch, 2012; Wicherts et al., 2012; Yarkoni, 2012; Zimmermann et al., 2011; Hartshorne and Schachner, 2012; Kravitz and Baker, 2011), but the need for an alternative system is clearly pressing (Casadevall and Fang, 2012). Given the data we surveyed above, almost anything appears superior to the status quo.

## Acknowledgements

Neil Saunders was of tremendous value in helping us obtain and understand the PubMed retraction data for Figure 1a. Ferric Fang and Arturo Casadeval were so kind as to let us use their retraction data to re-plot their figure on a logarithmic scale (Figure 1d). We are grateful to George A. Lozano, Vincent Larivière and Yves Gingras for sharing their citation data with us (Figure 3). We are indebted to John Ioannidis, Daniele Fanelli, Christopher Baker, Dwight Kravitz, Tom Hartley, Jason Priem, Stephen Curry, Nikolaus Kriegeskorte and four anonymous reviewers for their comments on an earlier version of this manuscript. MRM is a member of the UK Centre for Tobacco Control Studies, a UKCRC Public Health Research: Centre of Excellence. Funding from British Heart Foundation, Cancer Research UK, Economic and Social Research Council, Medical Research Council, and the National Institute for Health Research, under the auspices of the UK Clinical Research Collaboration, is gratefully acknowledged. BB was a Heisenberg-Fellow of the DFG during the time most of this manuscript was written and their support is gratefully acknowledged as well.

## References

Adam, D. (2002). The counting house. *Nature* 415, 726–9.

Adler, N. J., and Harzing, a.-W. (2009). When Knowledge Wins: Transcending the Sense and Nonsense of Academic Rankings. *Academy of Management Learning & Education* 8, 72–95.

Adler, R., Ewing, J., and Taylor, P. (2008). Joint Committee on Quantitative Assessment of Research: Citation Statistics (A report from the International Mathematical Union (IMU) in cooperation with the International Council of Industrial and Applied Mathematics (ICIAM) and the Institute of Mathemat. Available at: http://www.mathunion.org/fileadmin/IMU/Report/CitationStatistics.pdf.

Allen, L., Jones, C., Dolby, K., Lynn, D., and Walport, M. (2009). Looking for Landmarks: The Role of Expert Review and Bibliometric Analysis in Evaluating Scientific Publication Outputs. *PLoS ONE* 4, 8.

Anderson, M. S., Martinson, B. C., and De Vries, R. (2007). Normative dissonance in science: results from a national survey of u.s. Scientists. *Journal of empirical research on human research ethics : JERHRE* 2, 3–14.

Bachmann, T. (2011). Fair and Open Evaluation May Call for Temporarily Hidden Authorship, Caution When Counting the Votes, and Transparency of the Full Pre-publication Procedure. *Frontiers in computational neuroscience* 5, 61.

Bain, C. R., and Myles, P. S. (2005). Relationship between journal impact factor and levels of evidence in anaesthesia. *Anaesthesia and intensive care* 33, 567–70.

Baker, M. (2012). Independent labs to verify high-profile papers. *Nature.* Available at: http://www.nature.com/doifinder/10.1038/nature.2012.11176 [Accessed January 8, 2013].

Bauer, H. H. (2004). Science in the 21st Century : Knowledge Monopolies and Research Cartels. *Jour. Scient. Explor.* 18, 643–660.

Baylis, M., Gravenor, M., and Kao, R. (1999). Sprucing up one's impact factor. *Nature* 401, 322.

Begley, C. G., and Ellis, L. M. (2012). Drug development: Raise standards for preclinical cancer research. *Nature* 483, 531–533.

Bertamini, M., and Munafo, M. R. (2012). Bite-Size Science and Its Undesired Side Effects. *Perspectives on Psychological Science* 7, 67–71.

Beverungen, A., Bohm, S., and Land, C. (2012). The poverty of journal publishing. *Organization* 19, 929–938.

Birukou, A., Wakeling, J. R., Bartolini, C., Casati, F., Marchese, M., Mirylenka, K., Osman, N., Ragone, A., Sierra, C., and Wassef, A. (2011). Alternatives to peer review: novel approaches for research evaluation. *Frontiers in computational neuroscience* 5, 56.

Brown, E. N., and Ramaswamy, S. (2007). Quality of protein crystal structures. *Acta crystallographica. Section D, Biological crystallography* 63, 941–50.

Brumback, R. A. (2012). "3 . . 2 . . 1 . . Impact [factor]: target [academic career] destroyed!": just another statistical casualty. *Journal of child neurology* 27, 1565–76.

Button, K. S., Ioannidis, J. P. A., Mokrysz, C., Nosek, B. A., Flint, J., Robinson, E. S. J., and Munafò, M. R. (2013). Power failure: why small sample size undermines the reliability of neuroscience. *Nature Reviews Neuroscience* 14, 365–376.

Calcagno, V., Demoinet, E., Gollner, K., Guidi, L., Ruths, D., and De Mazancourt, C. (2012). Flows of Research Manuscripts Among Scientific Journals Reveal Hidden Submission Patterns. *Science (New York, N.Y.)* 338, 1065–1069.

Callaham, M. (2002). Journal Prestige, Publication Bias, and Other Characteristics Associated With Citation of Published Studies in Peer-Reviewed Journals. *JAMA: The Journal of the American Medical Association* 287, 2847–2850.

Carpenter, S. (2012). Psychology's Bold Initiative. *Science* 335, 1558–1561.

Casadevall, A., and Fang, F. C. (2012). Reforming science: methodological and cultural reforms. *Infection and immunity* 80, 891–6.

Chow, C. W., Haddad, K., Singh, G., and Wu, A. (2007). On Using Journal Rank to Proxy for an Article ' s Contribution or Value. *Issues in Accounting Education* 22, 411–427.

Cokol, M., Iossifov, I., Rodriguez-Esteban, R., and Rzhetsky, A. (2007). How many scientific papers should be retracted? *EMBO reports* 8, 422–3.

Collaboration, O. S. (2012). An Open, Large-Scale, Collaborative Effort to Estimate the Reproducibility of Psychological Science. *Perspectives on Psychological Science* 7, 657–660.

Curry, S. (2009). Eye-opening Access. *Occam's Typwriter: Reciprocal Space.* Available at: http://occamstypewriter.org/scurry/2009/03/27/eye_opening_access/.

Van Dongen, S. (2011). Associations between asymmetry and human attractiveness: Possible direct effects of asymmetry and signatures of publication bias. *Annals of human biology* 38, 317–23.

Dwan, K., Altman, D. G., Arnaiz, J. A., Bloom, J., Chan, A.-W., Cronin, E., Decullier, E., Easterbrook, P. J., Von Elm, E., Gamble, C., et al. (2008). Systematic review of the empirical evidence of study publication bias and outcome reporting bias. *PloS one* 3, e3081.

Editorial (2013). Beware the impact factor. *Nature materials* 12, 89.

Editorial (2005). Not-so-deep impact. *Nature* 435, 1003–1004.

Editorial (2012). The Well-Behaved Scientist. *Science* 335, 285–285.

EuropeanCommission (2010). Science and Technology Report.

Evangelou, E., Siontis, K. C., Pfeiffer, T., and Ioannidis, J. P. A. (2012). Perceived information gain from randomized trials correlates with publication in high-impact factor journals. *Journal of clinical epidemiology* 65, 1274–81.

Eve, M. P. (2012). Tear it down, build it up: the Research Output Team, or the library-as-publisher. *Insights: the UKSG journal* 25, 158–162.

Fanelli, D. (2010). Do pressures to publish increase scientists' bias? An empirical support from US States Data. *PloS one* 5, e10271.

Fanelli, D. (2009). How many scientists fabricate and falsify research? A systematic review and meta-analysis of survey data. *PloS one* 4, e5738.

Fanelli, D. (2011). Negative results are disappearing from most disciplines and countries. *Scientometrics* 90, 891–904.

Fang, F. C., and Casadevall, A. (2011). Retracted science and the retraction index. *Infection and immunity* 79, 3855–9.

Fang, F. C., Steen, R. G., and Casadevall, A. (2012). Misconduct accounts for the majority of retracted scientific publications. *Proceedings of the National Academy of Sciences of the United States of America* 109, 17028–33.

Finardi, U. (2013). Correlation between Journal Impact Factor and Citation Performance: An experimental study. *Journal of Informetrics* 7, 357–370.

Florian, R. V (2012). Aggregating post-publication peer reviews and ratings. *Frontiers in computational neuroscience* 6, 31.

Frank, M. (2003). Impact factors: arbiter of excellence? *Journal of the Medical Library Association : JMLA* 91, 4–6.

Garfield, E. (1999). Journal impact factor: a brief review. *CMAJ : Canadian Medical Association journal = journal de l'Association medicale canadienne* 161, 979–80.

Garwood, J. (2011). A conversation with Peter Lawrence, Cambridge. "The Heart of Research is Sick". *LabTimes* 2-2011, 24–31.

Gauchat, G. (2012). Politicization of Science in the Public Sphere: A Study of Public Trust in the United States, 1974 to 2010. *American Sociological Review* 77, 167–187.

Gauchat, G. (2010). The cultural authority of science: Public trust and acceptance of organized science. *Public Understanding of Science* 20, 751–770.

Ghosh, S. S., Klein, A., Avants, B., and Millman, K. J. (2012). Learning from open source software projects to improve scientific review. *Frontiers in computational neuroscience* 6, 18.

Giles, J. (2007). Breeding cheats. *Nature* 445, 242–3.

Giner-Sorolla, R. (2012). Science or Art? How Aesthetic Standards Grease the Way Through the Publication Bottleneck but Undermine Science. *Perspectives on Psychological Science* 7, 562–571.

Goldacre, B. (2011). I foresee that nobody will do anything about this problem – Bad Science. *Bad Science.* Available at: http://www.badscience.net/2011/04/i-foresee-that-nobody-will-do-anything-about-this-problem/ [Accessed March 8, 2012].

Gonon, F., Konsman, J.-P., Cohen, D., and Boraud, T. (2012). Why most biomedical findings Echoed by newspapers turn out to be false: the case of attention deficit hyperactivity disorder. *PloS one* 7, e44275.

Gordon, M. D. (1982). Citation ranking versus subjective evaluation in the determination of journal hierachies in the social sciences. *Journal of the American Society for Information Science* 33, 55–57.

Hamilton, J. (2011). Debunked Science: Studies Take Heat In 2011. *NPR.* Available at: http://www.npr.org/2011/12/29/144431640/debunked-science-studies-take-heat-in-2011 [Accessed March 8, 2012].

Hartshorne, J. K., and Schachner, A. (2012). Tracking Replicability as a Method of Post-Publication Open Evaluation. *Frontiers in Computational Neuroscience* 6, 8.

Hauser, J. R., and Katz, G. M. (1998). Metrics: you are what you measure! *European Management Journal* 16, 517–528.

Hegarty, P., and Walton, Z. (2012). The Consequences of Predicting Scientific Impact in Psychology Using Journal Impact Factors. *Perspectives on Psychological Science* 7, 72–78.

Hernán, M. A. (2009). Impact factor: a call to reason. *Epidemiology (Cambridge, Mass.)* 20, 317–8; discussion 319–20.

Hönekopp, J., and Khan, J. (2011). Future publication success in science is better predicted by traditional measures than by the h index. *Scientometrics* 90, 843–853.

House of Commons (2004). Scientific Publications: Free for all? *Tenth Report of Session 2003-2004,* vol II: Written evidence, Appendix 138. Available at: http://www.publications.parliament.uk/pa/cm200304/cmselect/cmsctech/399/399we163.htm [Accessed December 17, 2012].

Hunter, J. (2012). Post-publication peer review: opening up scientific conversation. *Frontiers in computational neuroscience* 6, 63.

Ietto-Gillies, G. (2012). The evaluation of research papers in the XXI century. The Open Peer Discussion system of the World Economics Association. *Frontiers in computational neuroscience* 6, 54.

Ioannidis, J. P. A. (2005a). Contradicted and initially stronger effects in highly cited clinical research. *JAMA : the journal of the American Medical Association* 294, 218–28.

Ioannidis, J. P. A. (2005b). Why most published research findings are false. *PLoS medicine* 2, e124.

Ioannidis, J. P. A. (2012). Why Science Is Not Necessarily Self-Correcting. *Perspectives on Psychological Science* 7, 645–654.

Ioannidis, J. P. A., and Panagiotou, O. A. (2011). Comparison of effect sizes associated with biomarkers reported in highly cited individual articles and in subsequent meta-analyses. *JAMA : the journal of the American Medical Association* 305, 2200–10.

Kelly, C. D. (2006). Replicating Empirical Research In Behavioral Ecology: How And Why It Should Be Done But Rarely Ever Is. *The Quarterly Review of Biology* 81, 221–236.

Kravitz, D. J., and Baker, C. I. (2011). Toward a new model of scientific publishing: discussion and a proposal. *Frontiers in computational neuroscience* 5, 55.

Kreiman, G., and Maunsell, J. H. R. (2011). Nine criteria for a measure of scientific output. *Frontiers in computational neuroscience* 5, 48.

Kriegeskorte, N. (2012). Open evaluation: a vision for entirely transparent post-publication peer review and rating for science. *Frontiers in computational neuroscience* 6, 79.

Kriegeskorte, N., Walther, A., and Deca, D. (2012). An emerging consensus for open evaluation: 18 visions for the future of scientific publishing. *Frontiers in computational neuroscience* 6, 94.

Kyrillidou, M., Morris, S., and Roebuck, G. (2012). ARL statistics. *American Research Libraries Digital Publications.* Available at: http://publications.arl.org/ARL_Statistics [Accessed March 18, 2012].

Lau, S. L., and Samman, N. (2007). Levels of evidence and journal impact factor in oral and maxillofacial surgery. *International journal of oral and maxillofacial surgery* 36, 1–5.

Lawrence, P. (2008). Lost in publication: how measurement harms science. *Ethics in Science and Environmental Politics* 8, 9–11.

Lawrence, P. A. (2002). Rank injustice. *Nature* 415, 835–6.

Lawrence, P. A. (2007). The mismeasurement of science. *Current biology : CB* 17, R583–5.

Lee, C. (2012). Open peer review by a selected-papers network. *Frontiers in computational neuroscience* 6, 1.

Lehrer, J. (2010). The decline effect and the scientific method. *New Yorker.* Available at: http://www.newyorker.com/reporting/2010/12/13/101213fa_fact_lehrer [Accessed March 8, 2012].

Lin, J. (2012). Cracking Open the Scientific Process: "Open Science" Challenges Journal Tradition With Web Collaboration. *New York Times.* Available at: http://www.nytimes.com/2012/01/17/science/open-science-challenges-journal-tradition-with-web-collaboration.html?pagewanted=all [Accessed March 8, 2012].

Liu, S. (2006). Top Journals' Top Retraction Rates. *Scientific Ethics* 1, 92–93.

Lopez-Cozar, E. D., and Cabezas-Clavijo, A. (2013). Ranking journals: Could Google Scholar Metrics be an alternative to Journal Citation Reports and Scimago Journal Rank? *ArXiv* 1303.5870, 26.

Lozano, G. A., Larivière, V., and Gingras, Y. (2012). The weakening relationship between the impact factor and papers' citations in the digital age. *Journal of the American Society for Information Science and Technology* 63, 2140–2145.

Lu, H., Zou, Q., Gu, H., Raichle, M. E., Stein, E. A., and Yang, Y. (2012). Rat brains also have a default mode network. *Proceedings of the National Academy of Sciences of the United States of America* 109, 3979–84.

Makel, M. C., Plucker, J. A., and Hegarty, B. (2012). Replications in Psychology Research: How Often Do They Really Occur? *Perspectives on Psychological Science* 7, 537–542.

Miller, K. R. (2012). America's Darwin Problem. *Huffington Post.* Available at: http://www.huffingtonpost.com/kenneth-r-miller/darwin-day-evolution_b_1269191.html [Accessed March 14, 2012].

Moed, H. F., and Van Leeuwen, T. N. (1996). Impact factors can mislead. *Nature* 381, 186.

Moed, H. F., and Van Leeuwen, T. N. (1995). Improving the accuracy of institute for scientific information's journal impact factors. *Journal of the American Society for Information Science* 46, 461–467.

Moed, H. F., Van Leeuwen, T. N., and Reedijk, J. (1996). A critical analysis of the journal impact factors ofAngewandte Chemie and the journal of the American

Chemical Society inaccuracies in published impact factors based on overall citations only. *Scientometrics* 37, 105–116.

Møller, A. P., and Jennions, M. D. (2001). Testing and adjusting for publication bias. *Trends in Ecology & Evolution* 16, 580–586.

Møller, A. P., Thornhill, R., and Gangestad, S. W. (2005). Direct and indirect tests for publication bias: asymmetry and sexual selection. *Animal Behaviour* 70, 497–506.

Munafò, M. R., Freimer, N. B., Ng, W., Ophoff, R., Veijola, J., Miettunen, J., Järvelin, M.-R., Taanila, A., and Flint, J. (2009). 5-HTTLPR genotype and anxiety-related personality traits: a meta-analysis and new data. *American journal of medical genetics. Part B, Neuropsychiatric genetics : the official publication of the International Society of Psychiatric Genetics* 150B, 271–81.

Munafò, M. R., Matheson, I. J., and Flint, J. (2007). Association of the DRD2 gene Taq1A polymorphism and alcoholism: a meta-analysis of case-control studies and evidence of publication bias. *Molecular psychiatry* 12, 454–61.

Nath, S. B., Marcus, S. C., and Druss, B. G. (2006). Retractions in the research literature: misconduct or mistakes? *The Medical journal of Australia* 185, 152–4.

Van Noorden, R. (2013). Mathematicians aim to take publishers out of publishing. *Nature.*

Van Noorden, R. (2011). Science publishing: The trouble with retractions. *Nature* 478, 26–8.

Nosek, B. A., and Bar-Anan, Y. (2012). Scientific Utopia: I. Opening Scientific Communication. *Psychological Inquiry* 23, 217–243.

Nosek, B. A., Spies, J. R., and Motyl, M. (2012). Scientific Utopia: II. Restructuring Incentives and Practices to Promote Truth Over Publishability. *Perspectives on Psychological Science* 7, 615–631.

Nowotny, H. (2005). Science and society. High- and low-cost realities for science and society. *Science (New York, N.Y.)* 308, 1117–8.

Obremskey, W. T., Pappas, N., Attallah-Wasif, E., Tornetta, P., and Bhandari, M. (2005). Level of evidence in orthopaedic journals. *The Journal of bone and joint surgery. American volume* 87, 2632–8.

Palmer, A. R. (2000). QUASI-REPLICATION AND THE CONTRACT OF ERROR: Lessons from Sex Ratios, Heritabilities and Fluctuating Asymmetry. *Annual Review of Ecology and Systematics* 31, 441–480.

Popper, K. (1995). *In Search of a Better World: Lectures and Essays from Thirty Years.* Routledge; New edition edition (December 20, 1995).

Pöschl, U. (2012). Multi-stage open peer review: scientific evaluation integrating the strengths of traditional peer review with the virtues of transparency and self-regulation. *Frontiers in computational neuroscience* 6, 33.

Priem, J. (2013). Scholarship: Beyond the paper. *Nature* 495, 437–440.

Priem, J., and Hemminger, B. M. (2012). Decoupling the scholarly journal. *Frontiers in computational neuroscience* 6, 19.

Priem, J., Piwowar, H. A., and Hemminger, B. M. (2012). Altmetrics in the wild: Using social media to explore scholarly impact. *ArXiv* 1203.4745.

Prinz, F., Schlange, T., and Asadullah, K. (2011). Believe it or not: how much can we rely on published data on potential drug targets? *Nature reviews. Drug discovery* 10, 712.

Rafols, I., Leydesdorff, L., O'Hare, A., Nightingale, P., and Stirling, A. (2012). How journal rankings can suppress interdisciplinary research: A comparison between Innovation Studies and Business & Management. *Research Policy* 41, 1262–1282.

Reedijk, J. (1998). Sense and nonsense of science citation analyses: comments on the monopoly position of ISI and citation inaccuracies. Risks of possible misuse and biased citation and impact data. *New Journal of Chemistry* 22, 767–770.

Research Information Network (2008). Activities, costs and funding flows in the scholarly communications system | Research Information Network. *Report commissioned by*

*the Research Information Network (RIN).* Available at: http://www.rin.ac.uk/our-work/communicating-and-disseminating-research/activities-costs-and-funding-flows-scholarly-commu [Accessed March 18, 2013].

Roemer, R. C., and Borchardt, R. (2012). From bibliometrics to altmetrics: A changing scholarly landscape. *College & Research Libraries News* 73, 596–600.

Rossner, M., Van Epps, H., and Hill, E. (2007). Show me the data. *The Journal of Cell Biology* 179, 1091–1092.

Saha, S., Saint, S., and Christakis, D. A. (2003). Impact factor: a valid measure of journal quality? *Journal of the Medical Library Association : JMLA* 91, 42–6.

Sandewall, E. (2012). Maintaining live discussion in two-stage open peer review. *Frontiers in computational neuroscience* 6, 9.

Sarewitz, D. (2012). Beware the creeping cracks of bias. *Nature* 485, 149–149.

Sarewitz, D. (2013). Science must be seen to bridge the political divide. *Nature* 493, 7.

Schooler, J. (2011). Unpublished results hide the decline effect. *Nature* 470, 437.

Scott, S., Kranz, J. E., Cole, J., Lincecum, J. M., Thompson, K., Kelly, N., Bostrom, A., Theodoss, J., Al-Nakhala, B. M., Vieira, F. G., et al. (2008). Design, power, and interpretation of studies in the standard murine model of ALS. *Amyotrophic lateral sclerosis : official publication of the World Federation of Neurology Research Group on Motor Neuron Diseases* 9, 4–15.

Seglen, P. O. (1994). Causal relationship between article citedness and journal impact. *Journal of the American Society for Information Science* 45, 1–11.

Seglen, P. O. (1992). The skewness of science. *Journal of the American Society for Information Science* 43, 628–638.

Seglen, P. O. (1997). Why the impact factor of journals should not be used for evaluating research. *BMJ* 314.

Shapin, S. (2008). *The scientific life : a moral history of a late modern vocation.* Chicago: University of Chicago Press.

Simmons, J. P., Nelson, L. D., and Simonsohn, U. (2011). False-positive psychology: undisclosed flexibility in data collection and analysis allows presenting anything as significant. *Psychological science* 22, 1359–66.

Simmons, L. W., Tomkins, J. L., Kotiaho, J. S., and Hunt, J. (1999). Fluctuating paradigm. *Proceedings of the Royal Society B: Biological Sciences* 266, 593–595.

Singh, G., Haddad, K. M., and Chow, C. W. (2007). Are Articles in "Top" Management Journals Necessarily of Higher Quality? *Journal of Management Inquiry* 16, 319–331.

Siontis, K. C. M., Evangelou, E., and Ioannidis, J. P. A. (2011). Magnitude of effects in clinical trials published in high-impact general medical journals. *International journal of epidemiology* 40, 1280–91.

Sønderstrup-Andersen, E. M., and Sønderstrup-Andersen, H. H. K. (2008). An investigation into diabetes researcher's perceptions of the Journal Impact Factor — reconsidering evaluating research. *Scientometrics* 76, 391–406.

Song, F., Eastwood, A., Gilbody, S., and Duley, L. (1999). The role of electronic journals in reducing publication bias. *Informatics for Health and Social Care* 24, 223–229.

Starbuck, W. H. (2005). How Much Better Are the Most-Prestigious Journals? The Statistics of Academic Publication. *Organization Science* 16, 180–200.

Statzner, B., and Resh, V. H. (2010). Negative changes in the scientific publication process in ecology: potential causes and consequences. *Freshwater Biology* 55, 2639–2653.

Steen, R. G. (2011a). Retractions in the scientific literature: do authors deliberately commit research fraud? *Journal of medical ethics* 37, 113–7.

Steen, R. G. (2011b). Retractions in the scientific literature: is the incidence of research fraud increasing? *Journal of medical ethics* 37, 249–53.

Sutton, J. (2011). psi study highlights replication problems. *The Psychologist News.* Available at:

http://www.thepsychologist.org.uk/blog/blogpost.cfm?threadid=1984&catid=48 [Accessed March 8, 2012].

Taylor, M., Perakakis, P., and Trachana, V. (2008). The siege of science. *Ethics in Science and Environmental Politics* 8, 17–40.

The PLoS Medicine Editors (2006). The impact factor game. It is time to find a better way to assess the scientific literature. *PLoS medicine* 3, e291.

Todd, P., and Ladle, R. (2008). Hidden dangers of a 'citation culture'. *Ethics in Science and Environmental Politics* 8, 13–16.

Tressoldi, P. E., Giofré, D., Sella, F., and Cumming, G. (2013). High Impact = High Statistical Standards? Not Necessarily So. *PloS one* 8, e56180.

Tsikliras, A. (2008). Chasing after the high impact. *Ethics in Science and Environmental Politics* 8, 45–47.

Upadhyay, J., Baker, S. J., Chandran, P., Miller, L., Lee, Y., Marek, G. J., Sakoglu, U., Chin, C.-L., Luo, F., Fox, G. B., et al. (2011). Default-mode-like network activation in awake rodents. *PloS one* 6, e27839.

Vanclay, J. K. (2011). Impact factor: outdated artefact or stepping-stone to journal certification? *Scientometrics* 92, 211–238.

Wager, E., and Williams, P. (2011). Why and how do journals retract articles? An analysis of Medline retractions 1988-2008. *Journal of medical ethics* 37, 567–70.

Wagner, P. D. (2011). What's in a number? *Journal of applied physiology (Bethesda, Md. : 1985)* 111, 951–3.

Walther, A., and Van den Bosch, J. J. F. (2012). FOSE: a framework for open science evaluation. *Frontiers in computational neuroscience* 6, 32.

Weale, A. R., Bailey, M., and Lear, P. A. (2004). The level of non-citation of articles within a journal as a measure of quality: a comparison to the impact factor. *BMC medical research methodology* 4, 14.

Welberg, L. (2012). Neuroimaging: Rats join the "default mode" club. *Nature Reviews Neuroscience* 11, 223.

Wicherts, J. M., Kievit, R. A., Bakker, M., and Borsboom, D. (2012). Letting the daylight in: Reviewing the reviewers and other ways to maximize transparency in science. *Frontiers in computational neuroscience* 6, 20.

Yarkoni, T. (2012). Designing next-generation platforms for evaluating scientific output: what scientists can learn from the social web. *Frontiers in computational neuroscience* 6, 72.

Yong, E. (2012). Replication studies: Bad copy. *Nature* 485, 298–300.

Young, N. S., Ioannidis, J. P. A., and Al-Ubaydli, O. (2008). Why current publication practices may distort science. *PLoS medicine* 5, e201.

Yue, W., Wilson, C. S., and Boller, F. (2007). Peer assessment of journal quality in clinical neurology. *Journal of the Medical Library Association : JMLA* 95, 70–6.

Zimmermann, J., Roebroeck, A., Uludag, K., Sack, A., Formisano, E., Jansma, B., De Weerd, P., and Goebel, R. (2011). Network-based statistics for a community driven transparent publication process. *Frontiers in computational neuroscience* 6, 11.

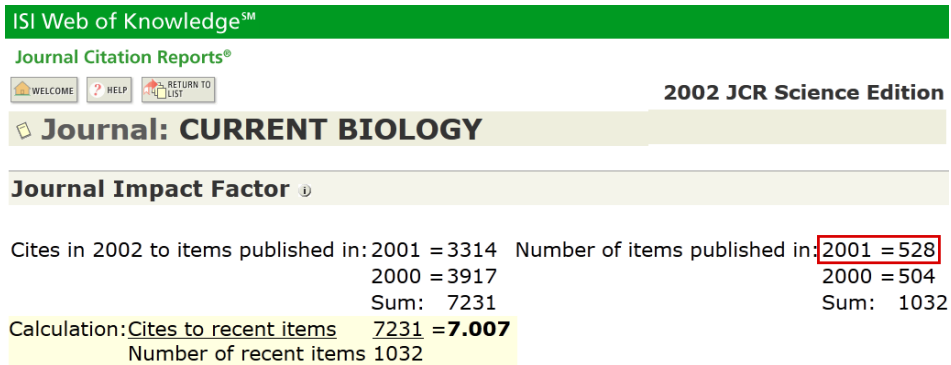

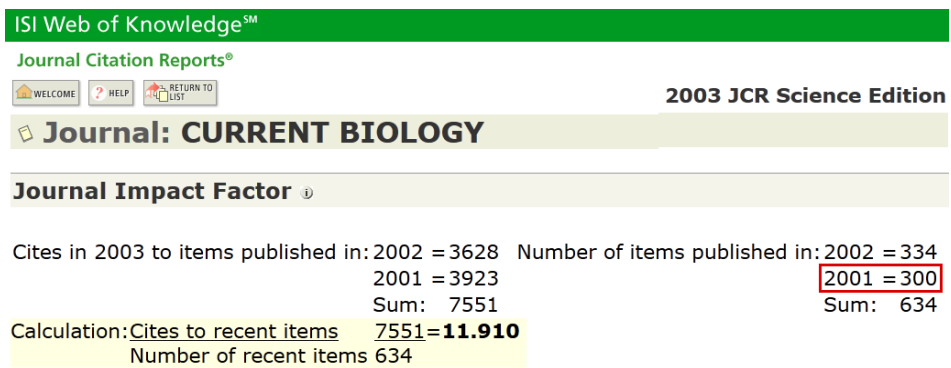

**Suppl. Fig. S1:** *Impact Factor of the journal „Current Biology" in the years 2002 (above) and 2003 (below) showing a 40% increase in impact.* The increase in the IF of the journal "Current Biology" from approx. 7 to almost 12 from one edition of Thomson Reuters' "Journal Citation Reports" to the next is due to a retrospective adjustment of the number of items published (marked), while the actual citations remained relatively constant.